\documentclass[12pt,a4paper,fleqn]{article}
\usepackage{epsfig}

\newcommand{\nn}{\noindent}
\newcommand{\tb}{\tan \beta}
\newcommand{\lesim}{\makebox[0pt][l]{\raisebox{2pt}{$\scriptstyle{<}$}}
\raisebox{-2pt}{$\scriptstyle{\sim}$}}
\newcommand{\gesim}{\makebox[0pt][l]{\raisebox{2pt}{$\scriptstyle{>}$}}
\raisebox{-2pt}{$\scriptstyle{\sim}$}}
\newcommand{\beq}{\begin{eqnarray}}
\newcommand{\eeq}{\end{eqnarray}}

\catcode`@=11
\def\citer{\@ifnextchar
[{\@tempswatrue\@citexr}{\@tempswafalse\@citexr[]}}

\def\@citexr[#1]#2{\if@filesw\immediate\write\@auxout{\string\citation{#2}}\fi
  \def\@citea{}\@cite{\@for\@citeb:=#2\do
    {\@citea\def\@citea{--\penalty\@m}\@ifundefined
       {b@\@citeb}{{\bf ?}\@warning
       {Citation `\@citeb' on page \thepage \space undefined}}%
\hbox{\csname b@\@citeb\endcsname}}}{#1}}
\catcode`@=12

\begin{document}

\begin{titlepage}
\vspace*{.1cm} 
\baselineskip=17pt

\nn PM/02--40 \hfill October 2002\\
PSI-PR-02-16 \\
\vspace*{0.9cm}

\begin{center}

{\large\sc {\bf MSSM Higgs particles in the intense--coupling regime}}\footnote{To appear in the proceedings of the 10th International Conference 
on Supersymmetry and Unification of Fundamental Interactions (SUSY02), June 
17-23, 2002, DESY Hamburg. Talk based on the collaboration with E.~Boos, 
A.~Djouadi and A.~Vologdin \cite{intense}.}

\vspace{0.7cm}

{\sc Margarete M.~M\"UHLLEITNER}\footnote{Present address: Paul Scherrer 
Institut, CH-5232 Villigen PSI, Switzerland; email: 
Margarete.Muehlleitner@psi.ch} 

\vspace*{6mm}

Laboratoire de Physique Math\'ematique et Th\'eorique, UMR5825--CNRS,\\
Universit\'e de Montpellier II, F--34095 Montpellier Cedex 5, France. 

\end{center} 

\vspace*{1cm} 

\begin{abstract}
\nn In the ``intense--coupling" regime all Higgs bosons of the 
Minimal Supersymmetric extension of the Standard Model (MSSM) are 
rather 
light and have comparable masses of ${\cal O}(100~{\rm GeV})$. They couple 
maximally to electroweak gauge bosons, and for large ratios of 
the vacuum expectation values of the two Higgs doublet fields, $\tb$, they 
interact strongly with the standard third generation fermions. We present in 
this note a comprehensive study of this scenario. We summarize the main 
phenomenological features, and the accordance with the direct constraints from 
Higgs boson searches at LEP2 and the Tevatron as well as the indirect 
constraints from precision measurements will 
be checked. After the presentation of the decay branching ratios, we 
discuss the production cross sections of the neutral Higgs particles in this 
regime at future colliders, the Tevatron Run II, the LHC and a 500 GeV 
$e^+e^-$ linear collider.
\end{abstract}
\end{titlepage}
%
%
%
%
\nn {\bf\large 1 Introduction}\\[0.2cm]
The Higgs sector of the MSSM consists of five physical Higgs states, two 
CP--even Higgs bosons, $h$ and $H$, a CP--odd $A$ and two charged $H^\pm$ 
bosons \cite{MSSM}. Supersymmetric relations imply that at least the lightest 
one, $h$, has a mass below $\sim 130$~GeV after including radiative 
corrections \cite{RC} and will therefore be accessible at future $pp$ 
and $e^+e^-$ colliders. 
In the decoupling regime \cite{decoup}, the $H,A$ and $H^\pm$ bosons are very 
heavy and degenerate, and the light $h$ state behaves Standard Model (SM) 
like. Being of ${\cal O}$(1~TeV) the heavy Higgs particles might escape 
detection so that the MSSM cannot be disentangled from the SM. A much more 
interesting scenario would be the opposite non--decoupling regime where all 
Higgs bosons are rather light and have comparable masses.
In the case of large values of $\tan\beta$ one of the CP--even bosons 
will be almost degenerate in mass with $A$ and have almost the same couplings 
as the pseudoscalar Higgs boson, while the other CP--even Higgs boson
behaves SM--like. In this situation all Higgs bosons will be accessible at the 
next generation of colliders in a plethora of production and decay processes 
with rates that can be very different from the SM case. Since the Higgs 
bosons are almost degenerate, several production channels have to be 
considered at the same time in order to detect the particles individually.

In the following, we discuss this intense--coupling regime 
taking into account direct and indirect experimental constraints. The various 
decay branching ratios will be presented, and the production at future 
colliders, the Tevatron Run II \cite{run2,cavalli}, the LHC \cite{cavalli,lhc} 
and $e^+e^-$ machines \cite{ee}, will be analyzed.\\[0.5cm]
%
%
%
%
\nn {\bf\large 2 The MSSM Higgs sector in the intense--coupling regime}\\[0.2cm]
The properties of the MSSM Higgs bosons are defined by their four masses, 
the ratio $\tan\beta$ and the mixing angle $\alpha$, introduced to diagonalize 
the Higgs boson mass matrix in the CP--even sector. Due to supersymmetry 
(SUSY), at tree level there are only two independent parameters, in general 
taken to be the $A$ boson mass, $M_A$, and $\tan\beta$. Radiative corrections 
introduce a further dependence on other soft SUSY breaking parameters 
\cite{RC}. The intense--coupling regime is characterized by rather light, 
almost degenerate Higgs bosons and large values of $\tan\beta$. Taking 
into account the leading radiative corrections in the case of large 
$\tan\beta$ for the Higgs boson masses and the couplings to 
fermions and gauge bosons, rather simple and accurate formulae can be derived, 
from which the features of the Higgs boson sector in this non-decoupling 
scenario can be read off \cite{intense}: The Higgs states are almost 
degenerate in mass, with  
\beq
90 \quad\lesim\quad M_\Phi &\lesim& 130 \quad \mathrm{GeV},\quad 
\Phi=h,H,A\nonumber \\
M_{H^\pm} &\lesim& 150 \quad \mathrm{GeV}
\eeq
For large values of $\tan\beta$ one of the CP--even Higgs bosons, denoted by 
$\Phi_A$ in the following, is degenerate in mass with $A$ and has large 
couplings to $b$ quarks and $\tau$ leptons, {\it i.e.} it behaves like a 
$A$ boson, whereas the other CP--even Higgs boson, denoted by $\Phi_H$, 
behaves SM--like with maximal couplings to gauge bosons and strong couplings 
to top quarks. Depending on whether the $A$ boson mass value is below or 
above a critical mass value $M_C$, the respective role is taken over by $h$ 
or $H$: 
\beq
M_A \ge M_C: \; M_H=M_A \quad \mathrm{and} \quad H=\Phi_A   
\quad\quad\;\; M_h=M_C \quad \mathrm{and} \quad h=\Phi_H \nonumber\\ 
M_A \le M_C: \; M_h=M_A \quad \mathrm{and} \quad h=\Phi_A  
\quad\quad\;\; M_H=M_C \quad \mathrm{and} \quad H=\Phi_H 
\label{icr}
\eeq
The critical mass $M_C$ is approximatively given by
\beq
M_C &=& \sqrt{M_Z^2+\epsilon} \qquad \mathrm{with} \\
\epsilon &=& \frac{3G_F}{\sqrt{2}\pi^2} \left\{ 
\frac{\overline{m}^4_t}{\sin^2\beta} 
\left[ t + \frac{X_t}{2} \right]- \left[ \frac{\overline{m}_t^2 M_Z^2 t}{2} +
\frac{2\alpha_s}{\pi} \frac{\overline{m}^4_t}{\sin^2\beta} 
(X_t t + t^2) \right] \right\}
\eeq
where
\beq
t=\log \frac{M_S^2}{m_t^2} \quad \mathrm{and} \quad 
X_t = \frac{2A_t^2}{M_S^2} \left(1-\frac{A_t^2}{12M_S^2} \right)
\eeq
$G_F$ denotes the Fermi constant, $\overline{m}_t$ the running 
$\overline{{\rm MS}}$ top quark mass at the scale $M_t = 175$~GeV, to account 
for the leading QCD 
corrections, $\alpha_s$ the strong coupling constant at $M_t$, $A_t$ the 
stop trilinear coupling and $M_S$ the common SUSY breaking mass scale.
For $M_A\gg M_C$, the decoupling limit is reached and $h$ behaves SM--like, 
cf.~(\ref{icr}).
\\[0.5cm]
%
%
%
%
\nn {\bf\large 3 Experimental constraints on the intense--coupling regime}
\\[0.2cm]
{\bf 3.1 Direct constraints}\smallskip \\
The search for Higgs bosons at LEP2 in the 
Higgs--strahlung process, $e^+e^- \to ZH^0$, has set a lower bound on the SM 
Higgs boson mass, $M_{H^0}> 114.4$~GeV at the 95\% confidence level 
\cite{smlim}. In the MSSM, this bound is valid for the lighter CP--even 
Higgs boson $h$ if its coupling to $Z$ bosons is SM--like, 
$g_{ZZh}^2/g_{ZZH^0}^2 \equiv \sin^2 (\beta-\alpha) \approx 1$, or for the 
heavier CP--even state $H$ in the case of a SM--like $ZZH$ coupling, 
$g_{ZZH}^2/g_{ZZH^0}^2 \equiv \cos^2 (\beta-\alpha) \approx 1$. 
\begin{figure}[ht]
\begin{center}
\epsfig{figure=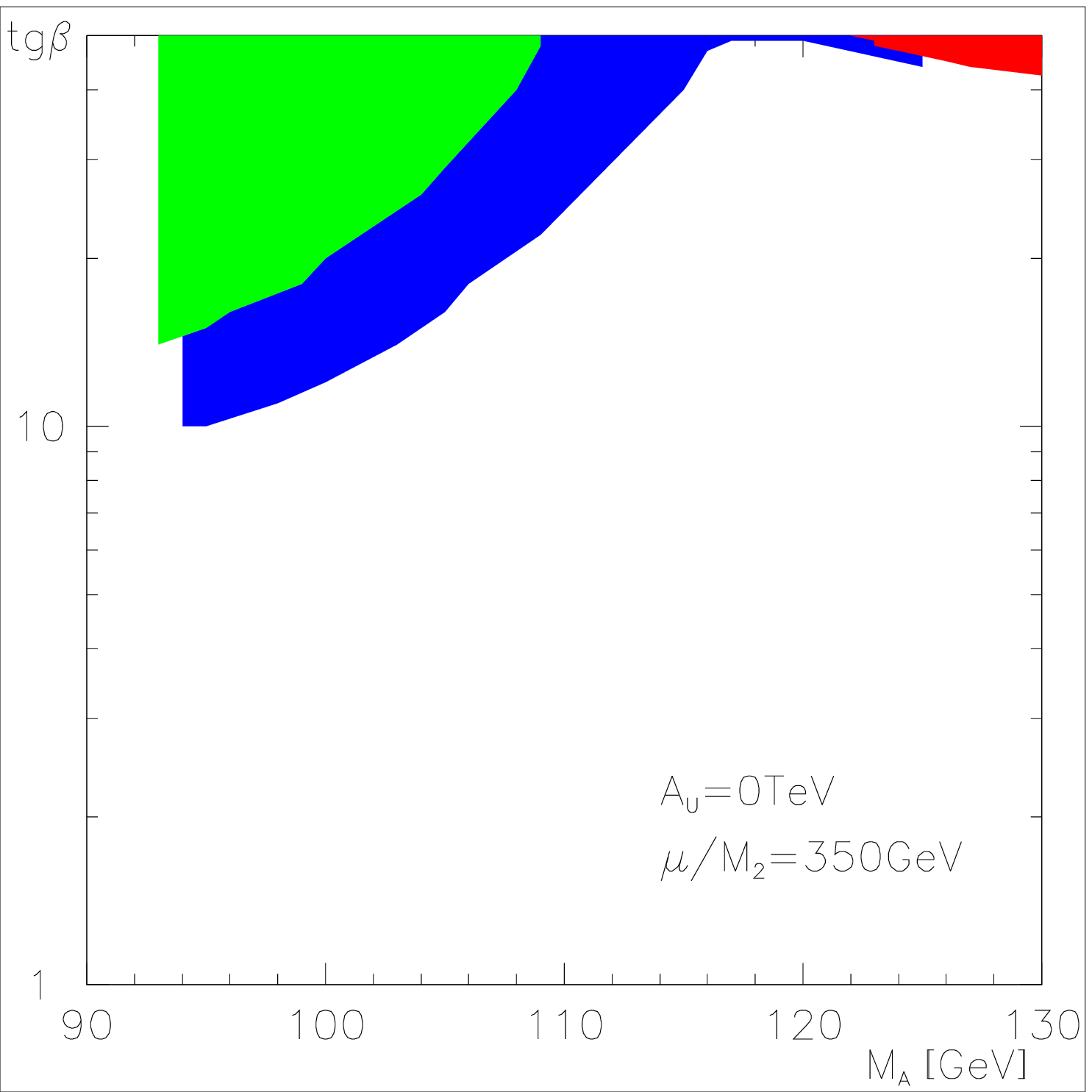,bbllx=3,bblly=3,bburx=410,bbury=420,width=5cm,clip=}
\hspace{1cm}
\epsfig{figure=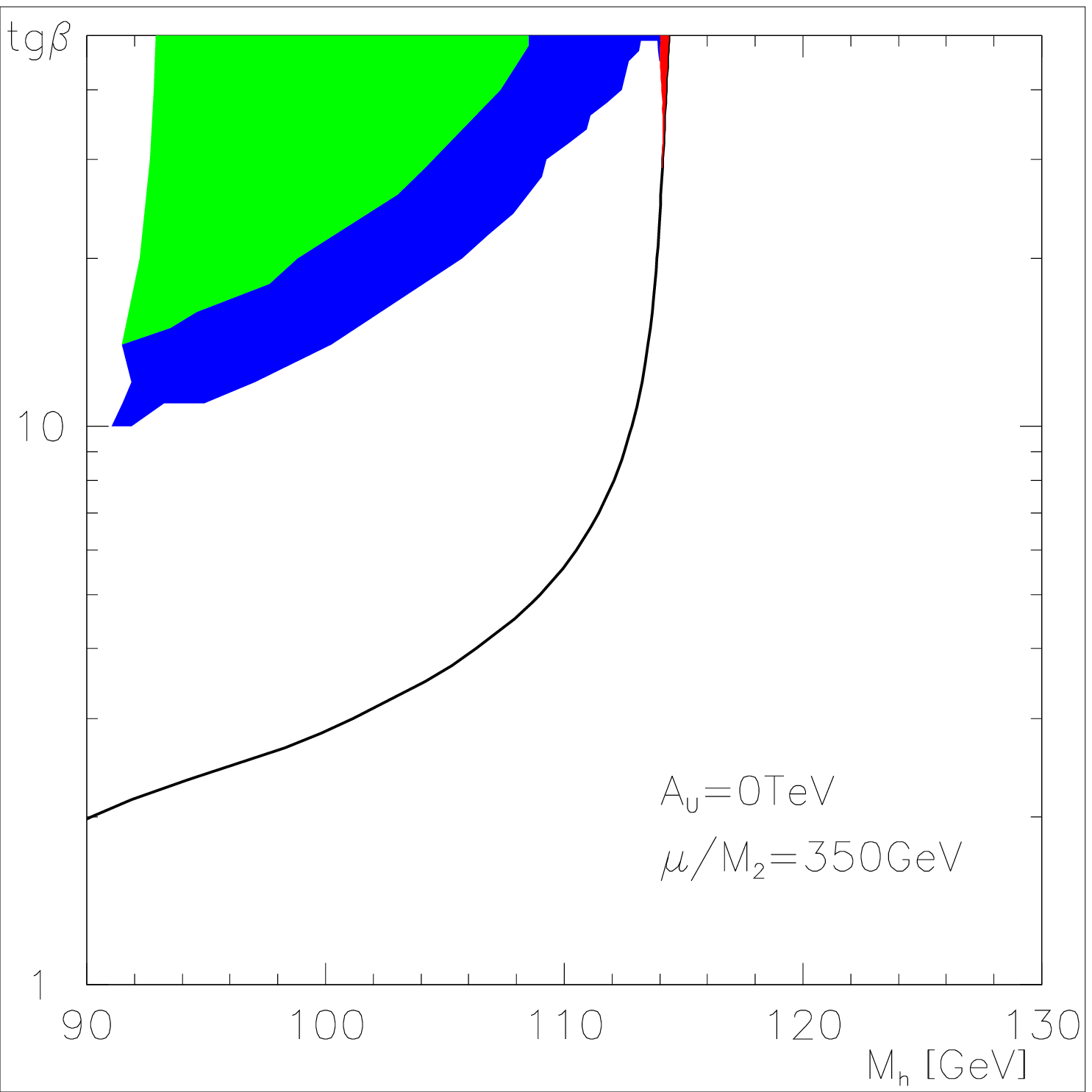,bbllx=3,bblly=3,bburx=410,bbury=420,width=5cm,clip=}
\\[0.2cm]
\epsfig{figure=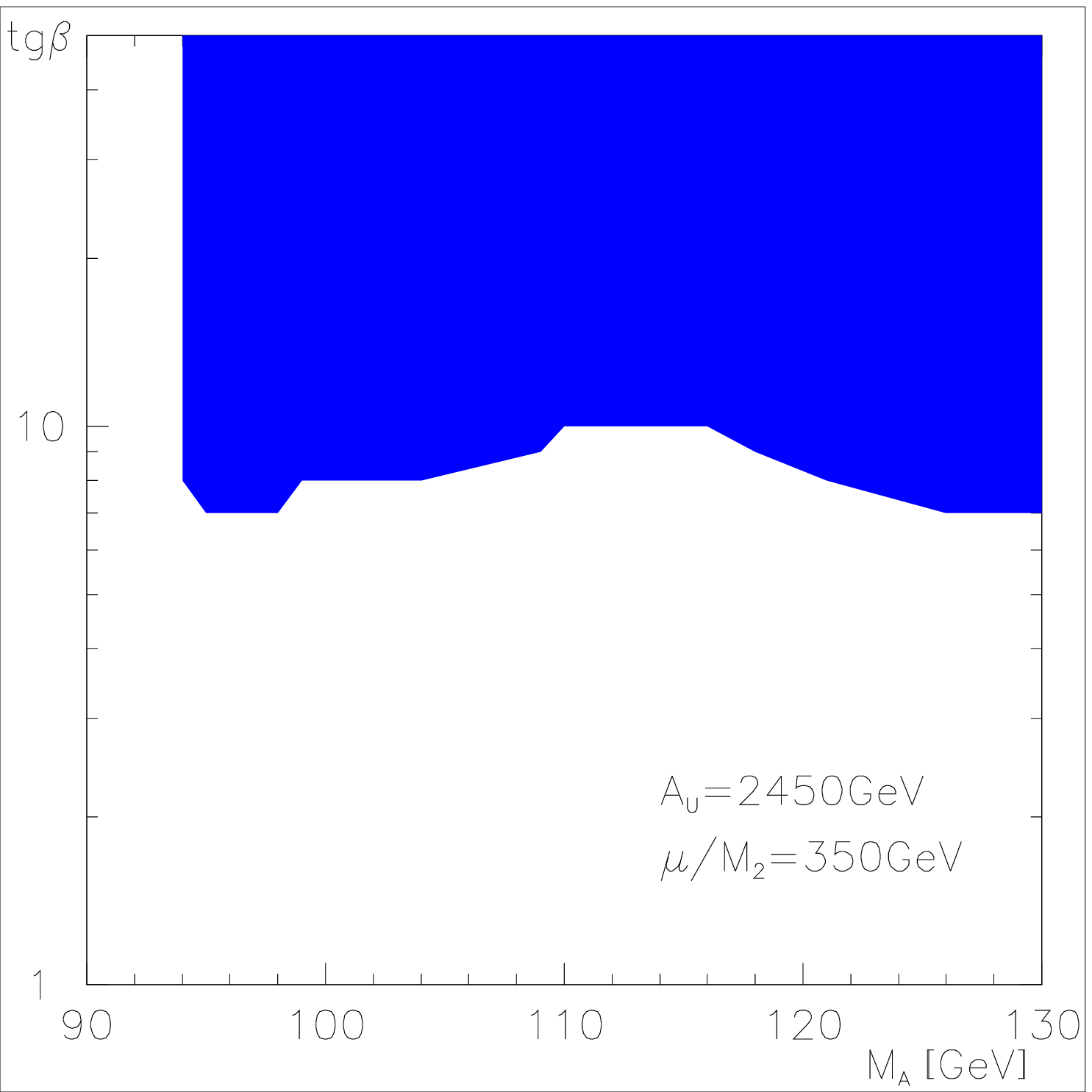,bbllx=3,bblly=3,bburx=410,bbury=420,width=5cm,clip=}
\hspace{1cm}
\epsfig{figure=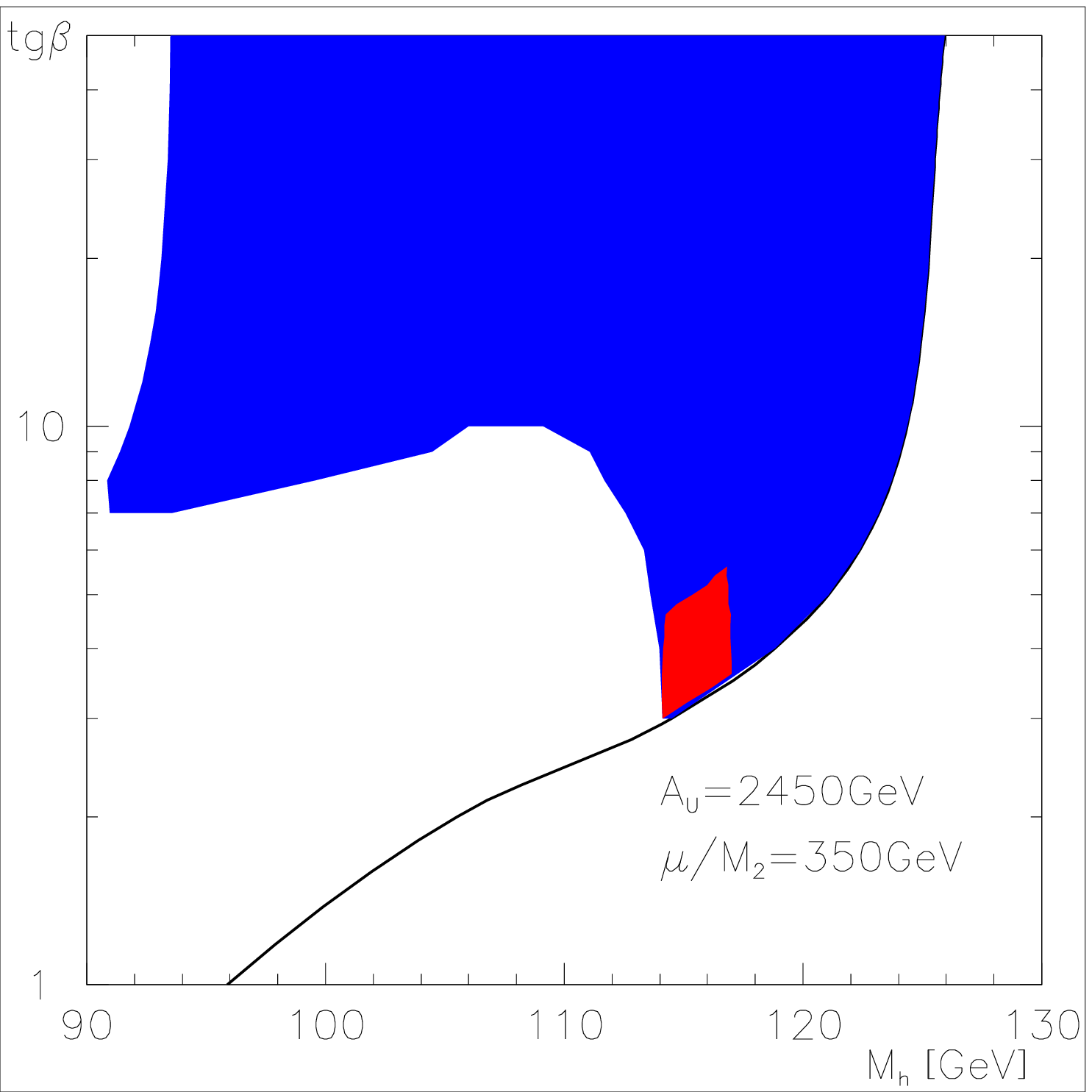,bbllx=3,bblly=3,bburx=410,bbury=420,width=5cm,clip=}
\caption{{\it The allowed regions for $M_A$ [left] and $M_h$ [right] from LEP2 searches as a function of $\tan\beta$ (colored regions) in the case of no mixing (up) and maximal mixing (down). The red regions indicate where $114$~GeV $<M_h<117$~GeV and $g_{ZZh}^2/g_{ZZH^0}^2>0.9$, and the green regions indicate where $114$~GeV $<M_H<117$~GeV and $g_{ZZH}^2/g_{ZZH^0}^2>0.9$.} 
}
\label{regions}
\end{center}
\vspace*{-.7cm}
\end{figure}
In addition, the search for Higgs bosons in the associated production process 
$e^+e^-\to Ah$ sets the limits $M_h > 91.0$~GeV and $M_A > 91.9$~GeV at the 
95\% confidence level \cite{smlim}. In order to derive a bound on $M_h$ for 
arbitrary values of $M_A$ and $\tan\beta$ we have fitted the exclusion plots 
for $\sin^2 (\beta-\alpha)$ and $\cos^2 (\beta-\alpha)$ versus $M_h$ and 
$M_A+M_h$, respectively, given in \cite{smlim}, and delineated the 
regions allowed by the LEP2 data up to $\sqrt{s}=209$~GeV in the 
[$M_A,\tan\beta$] and [$M_h,\tan\beta$] planes \cite{intense}. 
Fig.~\ref{regions} shows the domains allowed by the LEP2 constraints for the 
no--mixing scenario, {\it i.e.}~the trilinear soft SUSY breaking coupling in 
the stop sector $A_t=0$~TeV, and the maximal mixing scenario, $A_t \approx 
\sqrt{6}$~TeV. In the maximal mixing scenario, the allowed region is larger, 
since the large stop mixing increases the maximal $M_h$ value to the level 
where it exceeds the discovery limit, $M_h \;\gesim\; 114$~GeV. 

The red and the green regions show the implication of the 2.1$\sigma$ 
evidence for a SM--like Higgs boson with a mass $M_{\Phi_H}=115.6$~GeV 
\cite{smlim}. Considering the theoretical and experimental uncertainties, this 
result has been interpreted as favoring the mass range $114$~GeV 
$\lesim\; M_{\Phi_H}\;\lesim\; 117$~GeV. Since this Higgs boson should be 
SM--like, the additional constraint $g_{ZZ\Phi_H}^2/g_{ZZH^0}^2 \ge 0.9$ 
has been imposed. As can be seen in Fig.~\ref{regions}, the $H$ boson can be 
the ``observed'' Higgs boson only in the case of zero mixing. For large stop 
mixing, the $H$ boson mass always exceeds 117~GeV. 

Fig.~\ref{regions} demonstrates that the intense--coupling regime, where 
$90$~GeV $\lesim\; M_{\Phi}\;\lesim\; 130$~GeV [$\Phi=h,H,A$]
and $\tan\beta \gg 1$, is still allowed by the LEP2 searches.

\begin{figure}[htbp]
\begin{center}
\vspace{-0cm}
\epsfig{figure=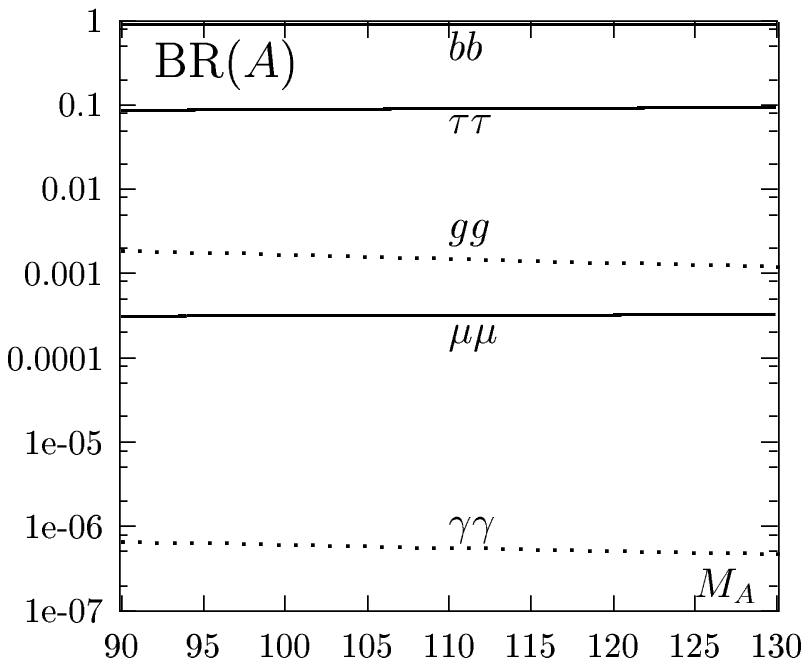,bbllx=142,bblly=574,bburx=375,bbury=766,height=4.3cm,width=4.2cm,clip=}
\hspace{0.5cm}
\epsfig{figure=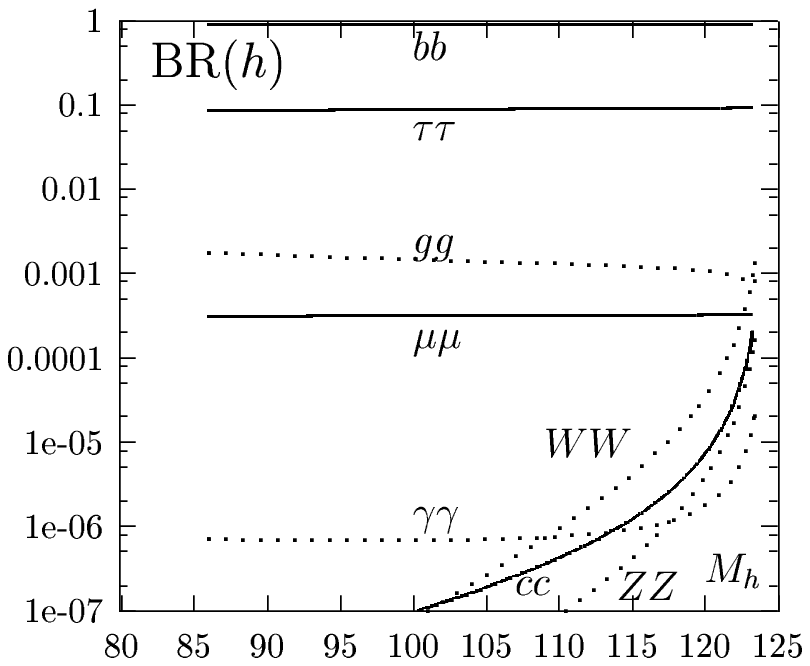,bbllx=142,bblly=574,bburx=375,bbury=766,height=4.3cm,width=4.2cm,clip=}
\hspace{0.5cm}
\epsfig{figure=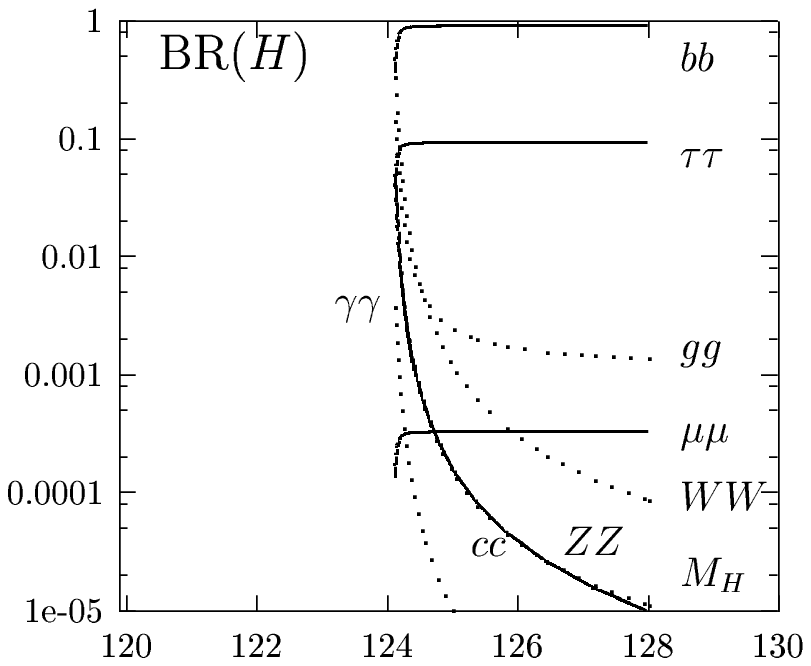,bbllx=142,bblly=574,bburx=375,bbury=766,height=4.3cm,width=4.2cm,clip=}
\caption{{\it The branching ratios of the neutral Higgs bosons $A$ (left), $h$ (middle), $H$ (right) as a function of their masses for $\tan\beta=30$ and $\mu=M_2=1$~TeV, $M_S=1~TeV$, $A_t=2.6$~TeV.}}
\label{bran}
\vspace*{-.9cm}
\end{center}
\end{figure}

\begin{figure}[htbp]
\begin{center}
\vspace{0cm}
\hspace{0.0cm}
\epsfig{figure=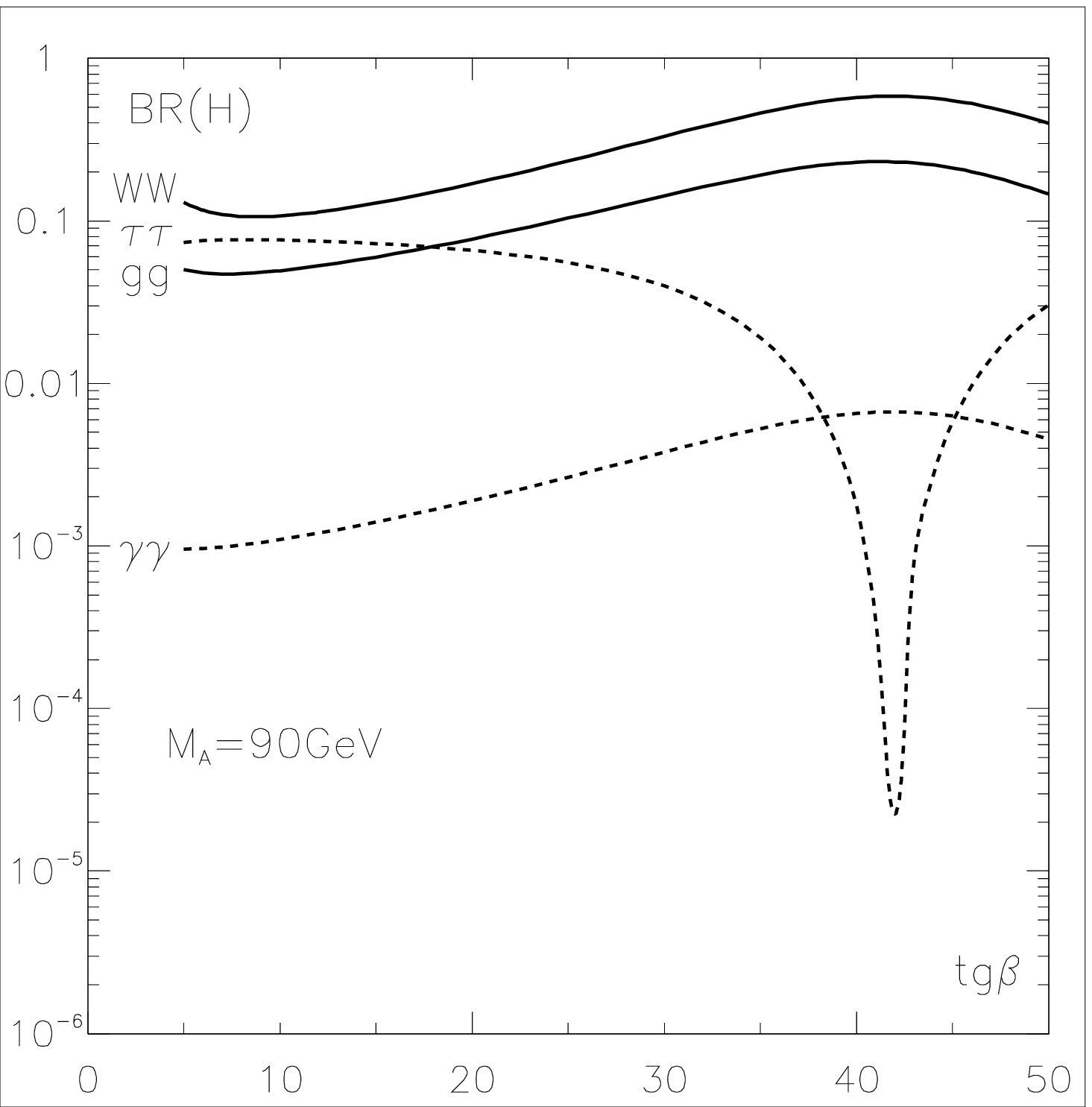,bbllx=5,bblly=1,bburx=415,bbury=422,height=4.2cm,width=4.2cm,clip=}\hspace{0.6cm}
\epsfig{figure=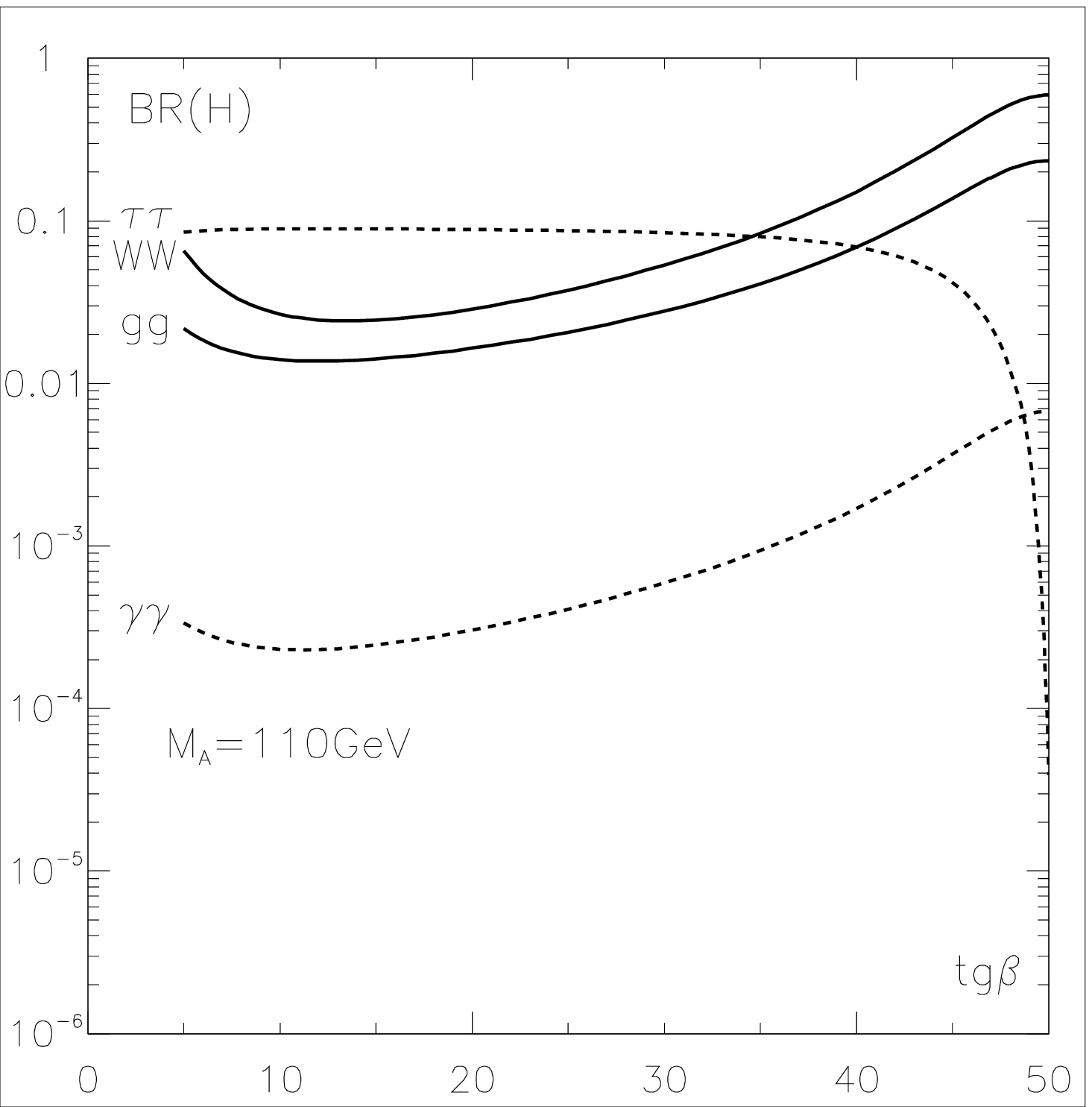,bbllx=5,bblly=1,bburx=415,bbury=422,height=4.2cm,width=4.2cm,clip=}\hspace{0.6cm}
\epsfig{figure=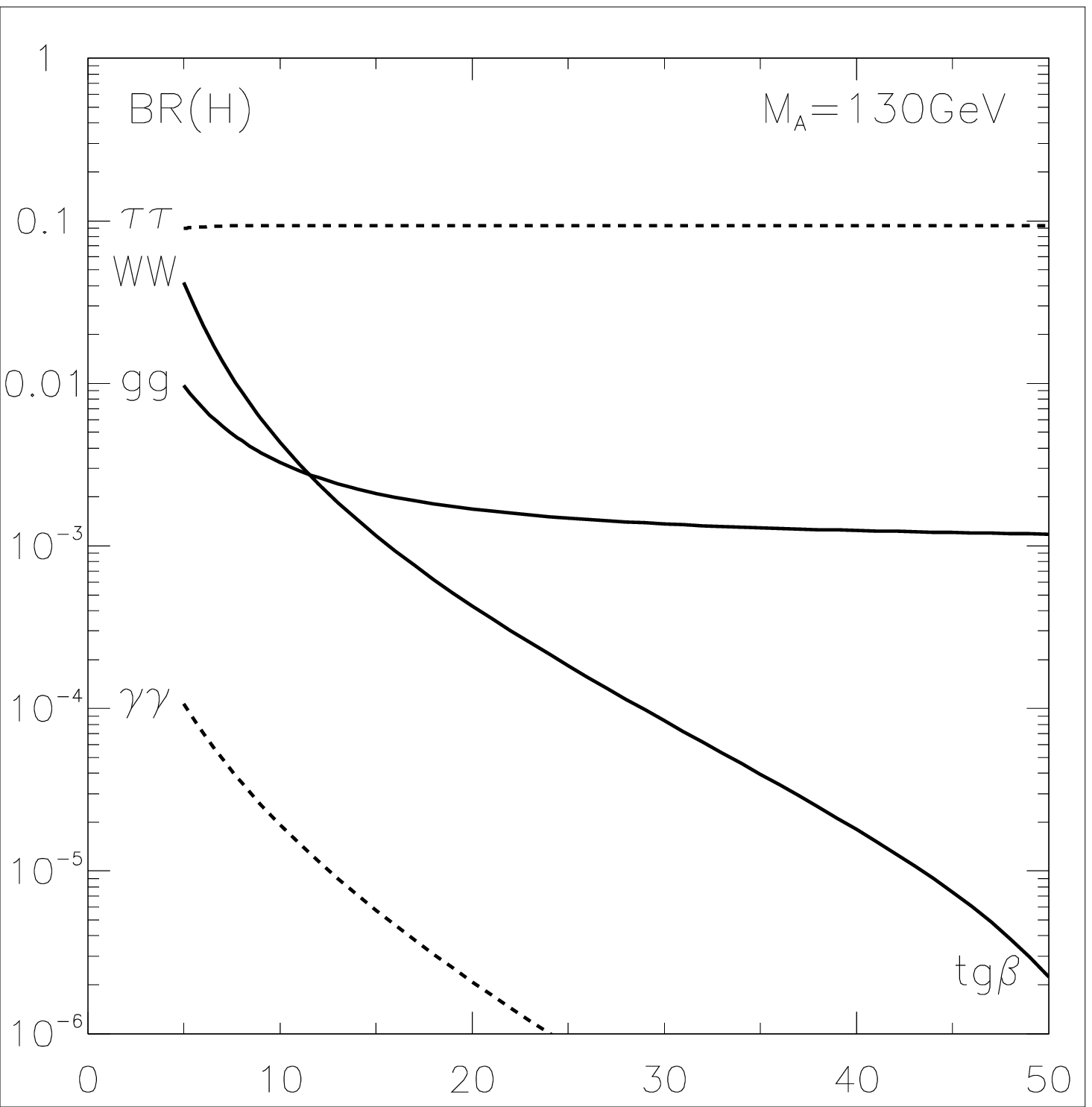,bbllx=5,bblly=1,bburx=415,bbury=422,height=4.2cm,width=4.2cm,clip=}
\vspace*{-0.1cm}
\caption{{\it The branching ratios of the $H$ boson into $\gamma\gamma$, $gg$, $W^+W^-$, $\tau^+\tau^-$ final states as a function of $\tan\beta$ for $M_A=90,110,130$~GeV.}}
\label{hbran}
\end{center}
\vspace*{-0.6cm}
\end{figure}

Further constraints can be obtained from the Higgs boson searches at the 
Tevatron \cite{cavalli}. The search for the top decay mode into a $b$-quark 
and a charged Higgs boson, $t \to b H^+$, sets the limits $\tan\beta\; 
\lesim \;50$ for $M_{H^\pm} \;\lesim\; 150$~GeV ({\it i.e.} $M_A\;\lesim\; 
130$~GeV). The data on $\tau^+\tau^- + 2$ jets can be exploited to derive 
limits from the process $pp \to b\bar{b}A$, leading to $\tan\beta\;\lesim\;80$ 
for $M_A\;\lesim\;130$~GeV. In summary, the intense--coupling regime is not 
excluded by the Tevatron data. \\[0.2cm]
{\bf 3.2 Indirect constraints}\smallskip \\
Indirect constraints on the parameters of the MSSM Higgs sector come from 
high precision data \cite{prec}. 
We examined the contributions of the MSSM Higgs sector in 
the intense--coupling regime \cite{intense} and found 
that they do not violate the 
experimental constraints on the $\rho$ parameter (the new physics 
contribution is limited to $\Delta_\rho^{NP} \;\lesim\; 1.1 \!\cdot\! 
10^{-3}$), the anomalous magnetic moment ($a_\mu=11659202(20) \!\cdot\! 
10^{-10}$), the $Zb\bar{b}$ vertex (the forward-backward asymmetry of the 
decay $Z\to b\bar{b}$ is $A^b_{FB}=0.099\pm0.002$) and the 
$b \to s\gamma$ decay (the branching ratio is given by $BR(b\to s\gamma) = (3.37\pm0.37\pm0.34\pm0.24^{+0.35}_{-0.16}\pm0.38)\!\cdot\! 10^{-4}$). 
\\[0.5cm]
%
%
%
%
\nn {\bf\large 4 Branching ratios and total widths}
\\[0.2cm]
Fig.~\ref{bran} shows the branching ratios of the neutral Higgs bosons $A,h,H$
as a function of their respective mass for 
$90\;\lesim\; M_A\;\lesim\; 130$~GeV and $\tan\beta =30$.  
As can be inferred from the figure, the $A$ boson decays with a ratio of 
90\% into $b\bar{b}$ and of 10\% into $\tau^+ \tau^-$, the other branching 
ratios being below $\sim 10^{-3}$. The pattern for the branching ratios of 
the CP--even Higgs bosons is similar to that of the $A$ boson if their masses 
are close to $M_A$ and $\tan\beta \gg 1$, unless $h$ and $H$ have masses 
close to $M_C$, where they behave SM--like. In practice, however, this limit 
is not reached, especially for $h$ with $\tan\beta\;\lesim\; 50$. As can be 
seen in Fig.~\ref{bran}, the $h$ branching ratios are all below $\sim 10^{-3}$ 
except for the $b\bar{b}$ and $\tau^+\tau^-$ final states. This is because 
the $h$ coupling to gauge bosons does not reach the SM limit, while the 
$hb\bar{b}$ coupling is still enhanced for large $\tan\beta$. In the case of 
$H$ and $M_H\;\gesim\; 125$~GeV the branching ratios into $\gamma\gamma$, $WW$ 
and $gg$ are much smaller than in the SM. They approach the SM values near 
the critical mass. For larger values of $\tan\beta$, however, a new feature 
occurs. The $H$ coupling to down type fermions becomes strongly suppressed 
and at some stage nearly vanishes, so that the branching ratios into 
$WW,gg$ and $\gamma\gamma$ become larger than in the SM case in this 
pathological region \cite{caretal}, cf.~Fig.~\ref{hbran}. 
\\
Finally, the total decay widths of $h,H$ are SM--like for masses near $M_C$ 
and therefore rather small. Otherwise, the total width of the $\Phi_A$--like 
boson, dominated by the decays in $b$ and $\tau$, being proportional to 
$\tan^2\beta$, can become rather large for large $\tan\beta$ values 
[{\it e.g.} $\Gamma_{tot}^H\approx 7$~GeV for $\tan\beta=50$, $M_A=130$~GeV], 
cf.~Fig.~4.\\[0.5cm] 
%
%
%
%
\nn {\bf\large 5 Production at Future Colliders}
\\[0.2cm]
{\bf 5.1 Production at $pp$ colliders.}\smallskip \\
The neutral MSSM Higgs bosons can be produced in a plethora of processes at
Tevatron 
\begin{picture}(30,130)
\put(60,10){\epsfig{figure=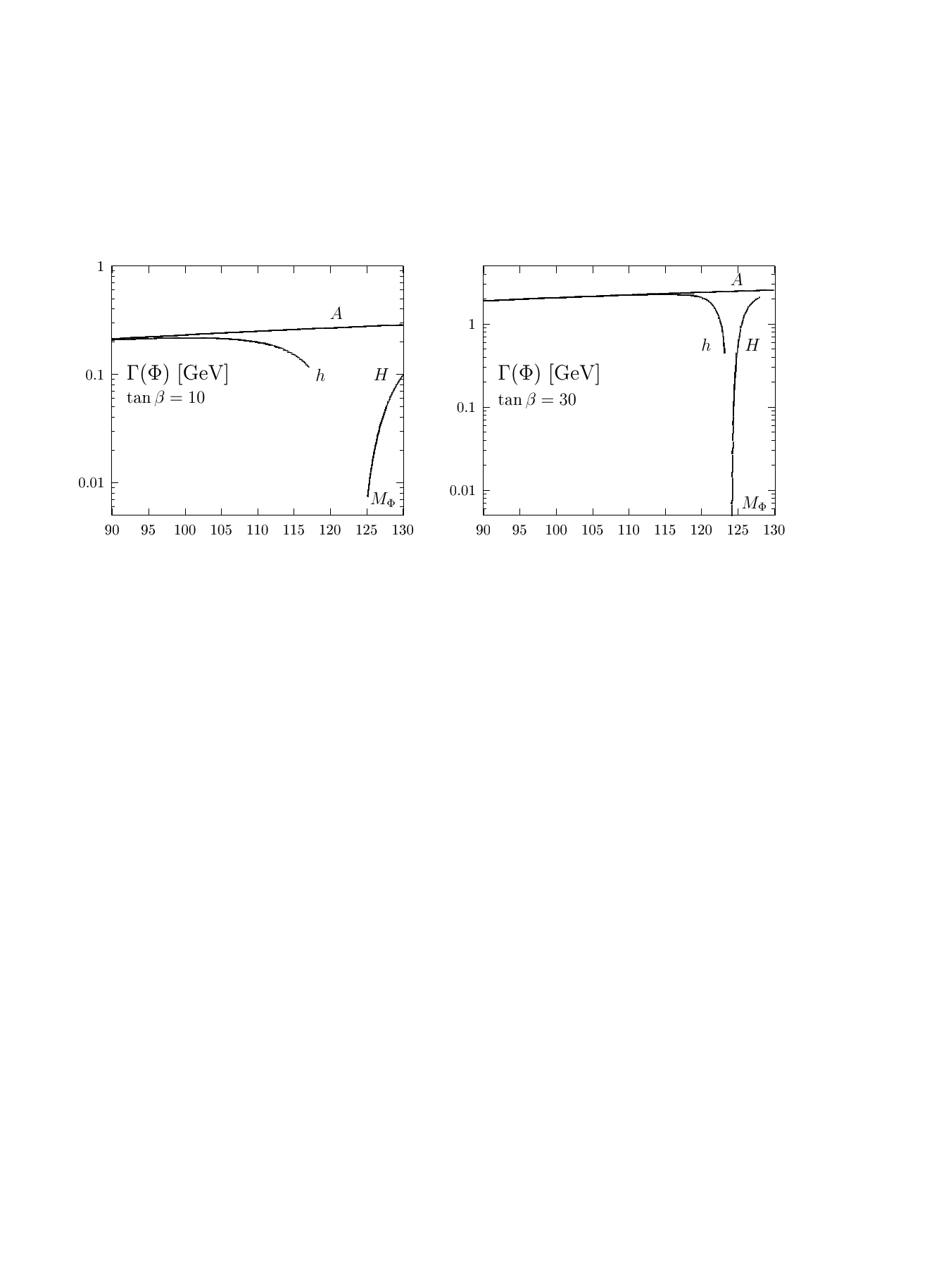,bbllx=305,bblly=511,bburx=538,bbury=699,height=4cm,width=4.3cm,clip=}}
\put(200,100){\parbox[t]{7cm}{{Figure 4: 
{\it The total decay widths of the neutral Higgs bosons $h,H,A$ as a function 
of their masses for $\tan\beta=30$.}
}}}
\label{tot}
\end{picture}\\[0cm]
Run II and the LHC. Fig.~\ref{lhcprod} shows the main production processes at 
the LHC for $\tan\beta=30$, {\it i.e.}
gluon--gluon fusion, $gg\to \Phi$ [$\Phi= h,H,A$], associated production, 
$gg,q\bar{q} \to \Phi + b\bar{b}/t\bar{t}$, and $WW/ZZ$ fusion, 
$VV\to h,H$ [$V=W,Z$].
For large $\tan\beta$ the $gg$ fusion process is dominated by the $b$ quark 
loop contributions, which are enhanced by $\tan^2\beta$ factors, leading to 
cross sections of up to 1500 pb for the $\Phi_A$--like bosons. Due to the 
large QCD backgrounds, only the $\gamma\gamma$ final states can be probed. 
\addtocounter{figure}{1}
\begin{figure}[htbp]
\begin{center}
\vspace{0.1cm}
\epsfig{figure=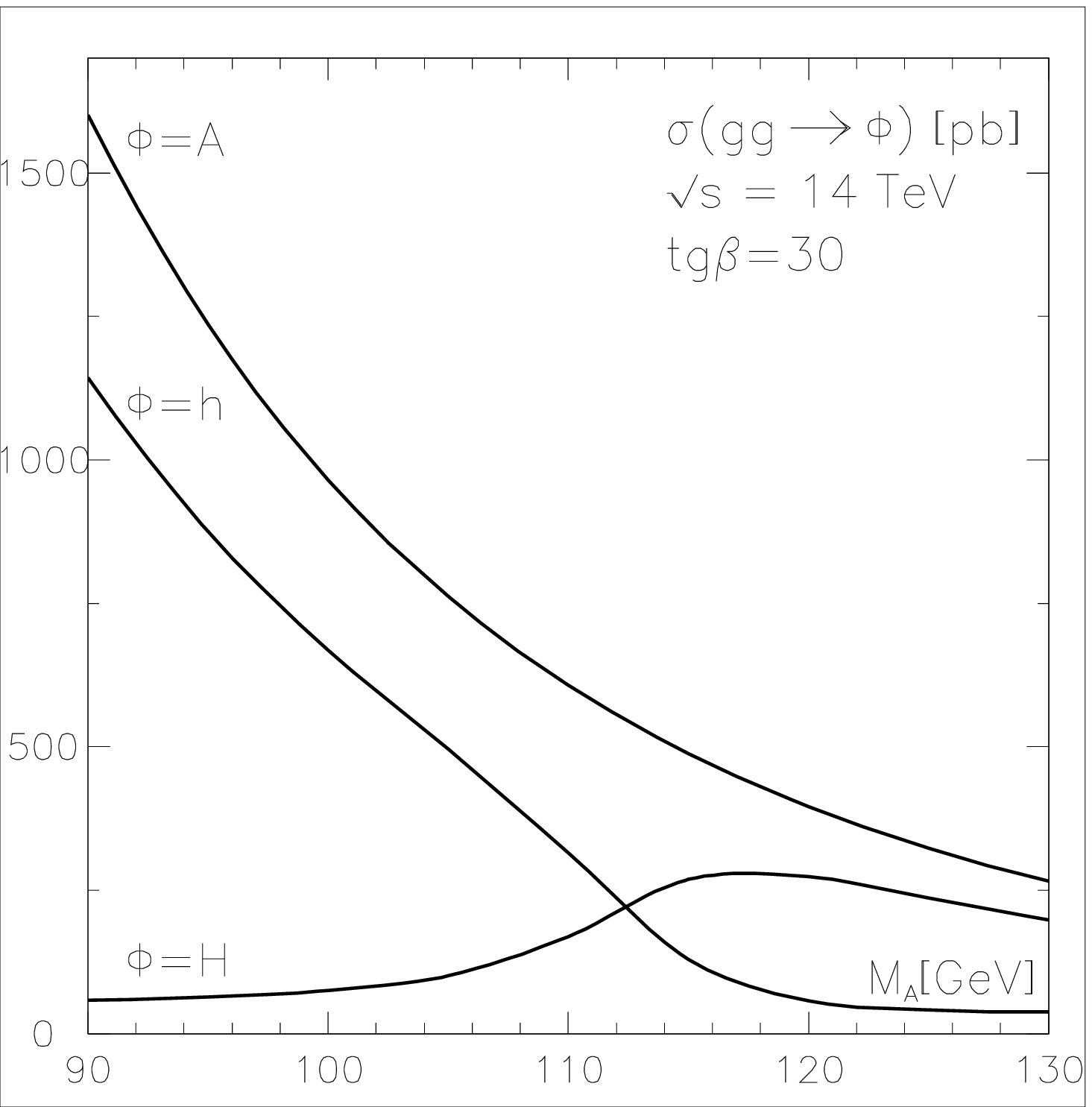,bbllx=20,bblly=26,bburx=410,bbury=408,height=3.5cm,width=3.5cm,clip=}\hspace{0.3cm}
\epsfig{figure=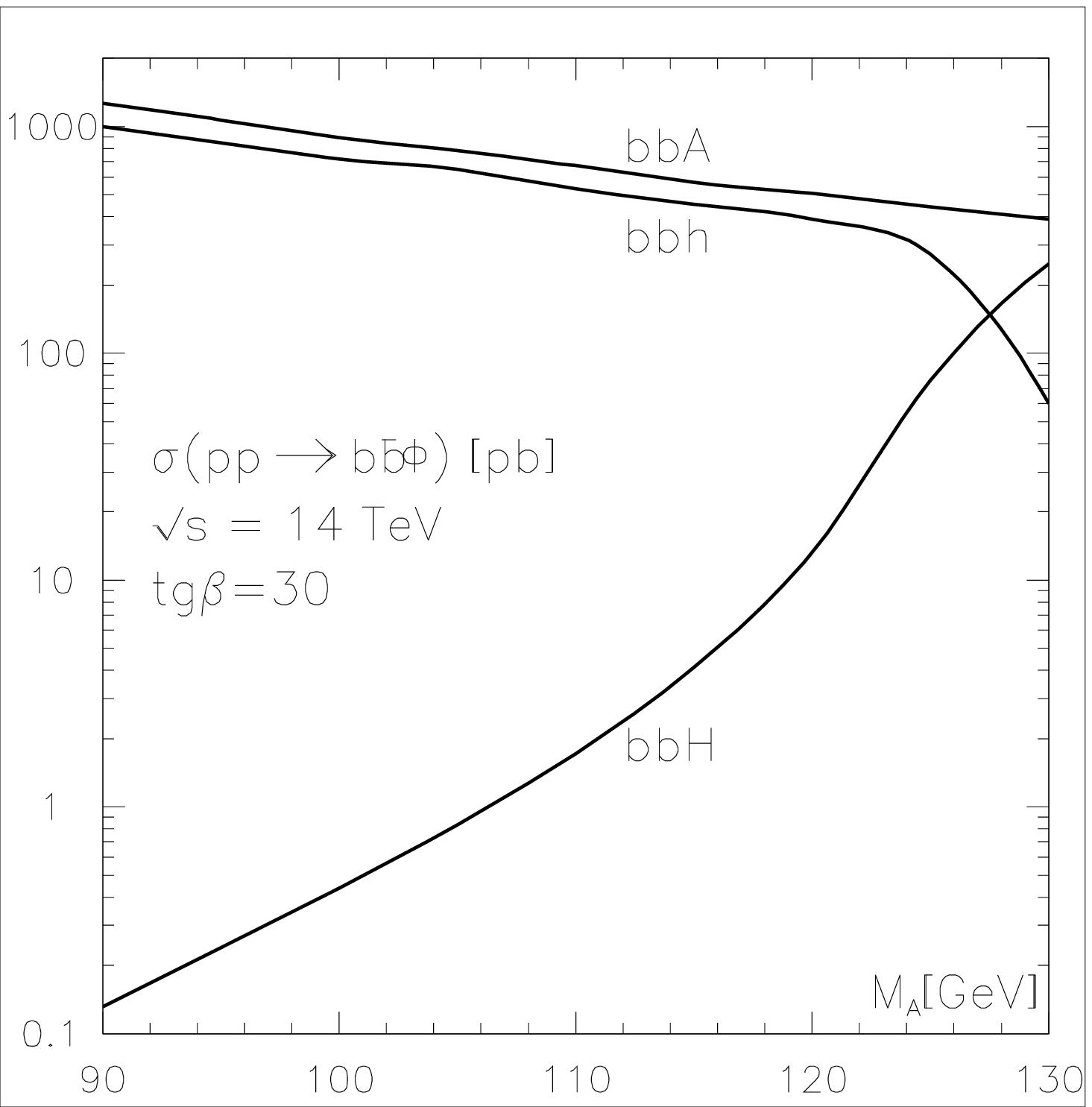,bbllx=20,bblly=26,bburx=410,bbury=408,height=3.5cm,width=3.5cm,clip=}\hspace{0.3cm}
\epsfig{figure=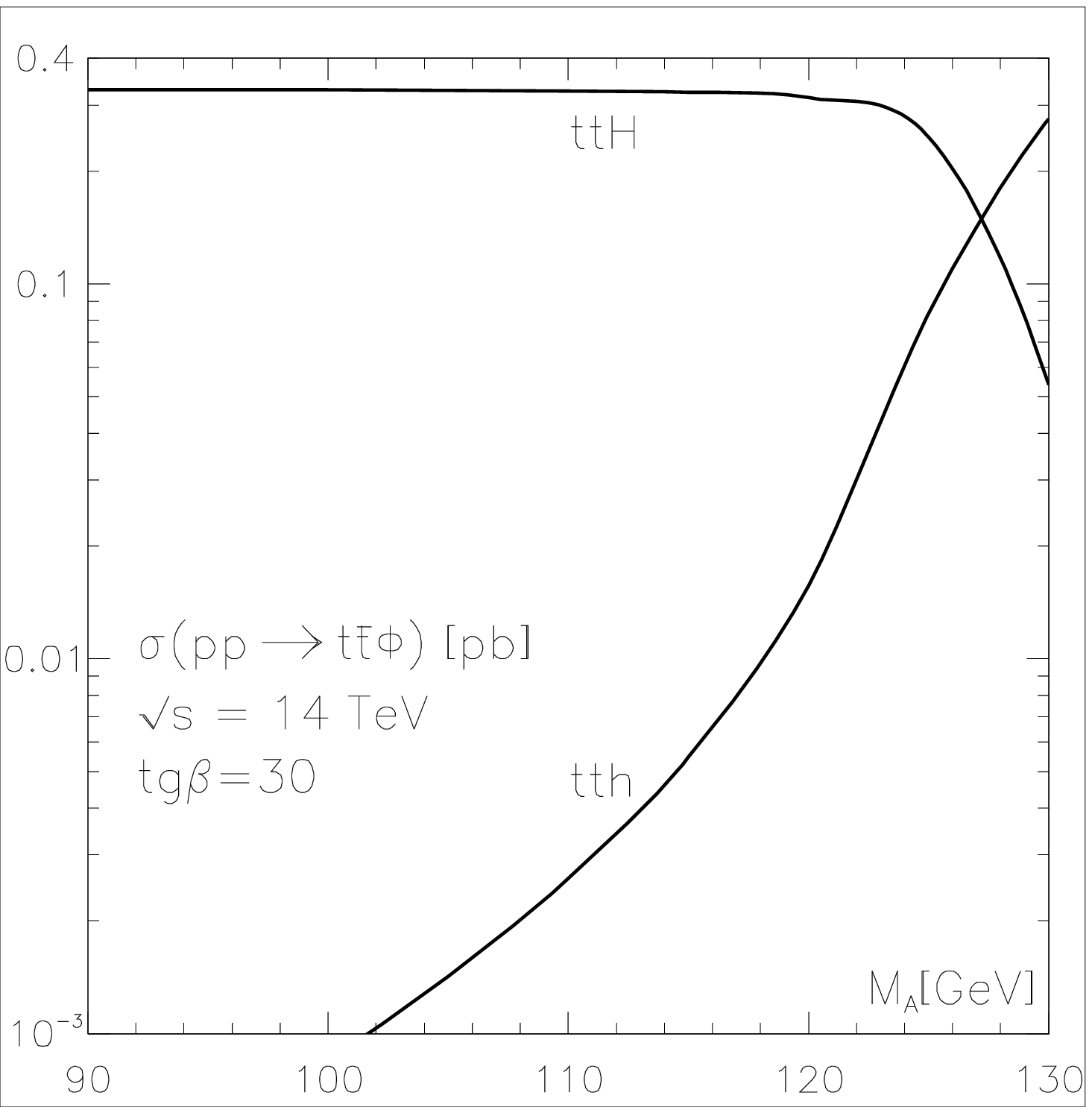,bbllx=20,bblly=26,bburx=410,bbury=408,height=3.5cm,width=3.5cm,clip=}\hspace{0.3cm}
\epsfig{figure=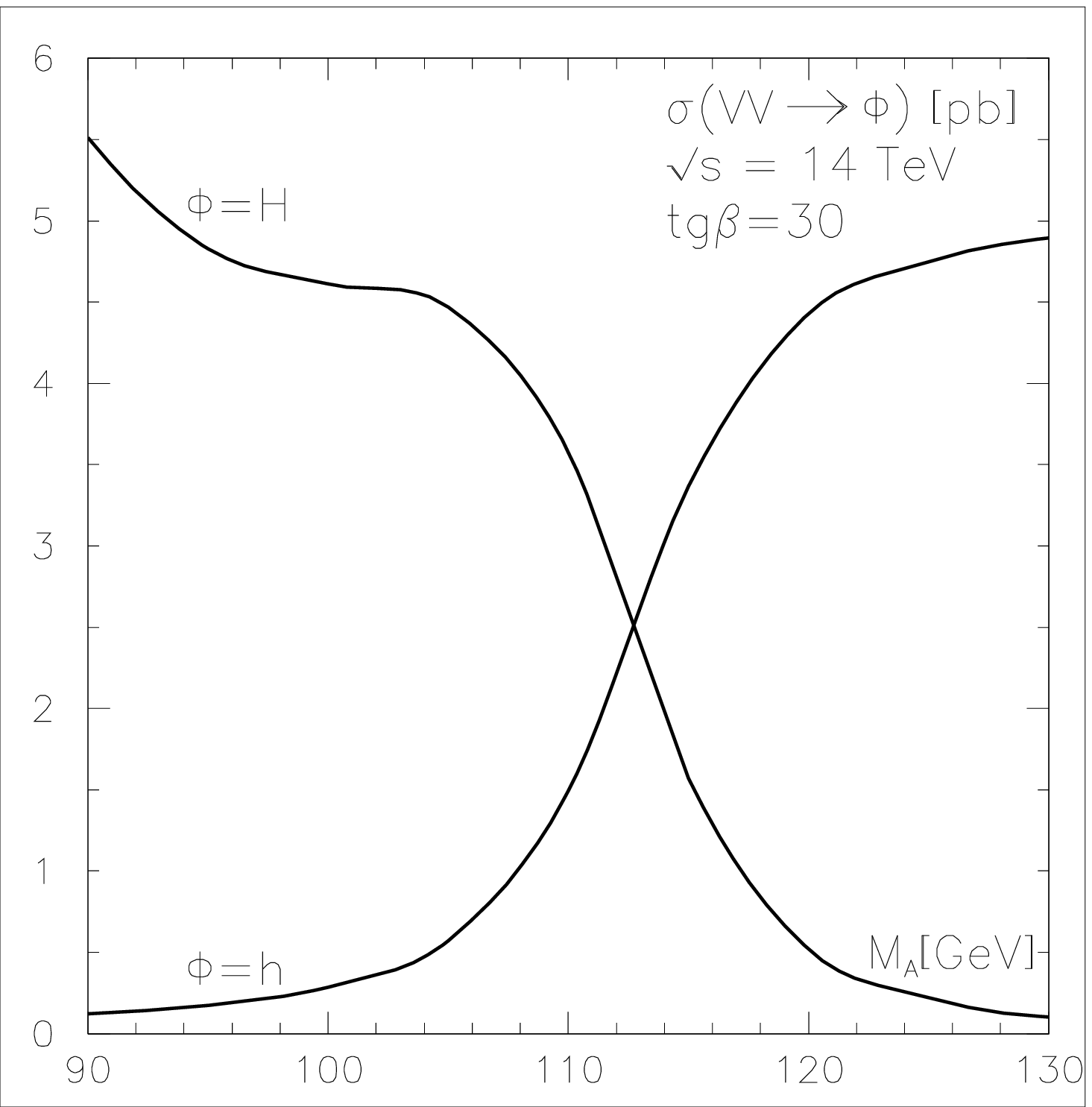,bbllx=20,bblly=26,bburx=410,bbury=408,height=3.5cm,width=3.5cm,clip=}\vspace*{-.2cm}
\end{center}
\caption{{\it Production processes at the LHC for the gluon-gluon fusion process, associated production with $b$ quarks and with top quarks, and the $WW/ZZ$ fusion processes (from left to right) as a function of $M_A$ for $\tan\beta=30$
with the same SUSY parameters as in Fig.~\ref{bran}. The next-to-leading order QCD corrections have been taken into account, where available \cite{spira}.}}
\label{lhcprod}
\vspace*{-.1cm}
\end{figure}
In the intense--coupling regime there can be situations, however, where the 
rates $\sigma (gg\to \Phi) \times BR(\Phi\to \gamma\gamma)$ are small for all 
three Higgs bosons, due to small cross sections, cf.~Fig.~\ref{lhcprod}, or 
small branching ratios into $\gamma\gamma$ in the case, where the SM limit is 
not yet reached, cf.~Fig.~\ref{bran}. The $A,\Phi_A$ production processes in 
association with $b$ quarks are strongly enhanced by $\tan^2\beta$ factors,
reaching rates similar to the $gg$ fusion process, whereas the 
$\Phi_H b\bar{b}$ cross section is much smaller due to a tiny Yukawa coupling. 
The detection of $A$ and $\Phi_A$, however, is promising, since the 
additional two $b$ quarks in the final state can be exploited to reduce the 
QCD backgrounds and one does not have to rely on the decays into photons or 
massive gauge bosons. The associated production cross sections with $t$ quark 
pairs are suppressed due to the smaller phase space and a not enhanced 
Higgs--$t\bar{t}$ coupling. To the contrary, the $A/\Phi_A t\bar{t}$ process 
is 
strongly suppressed by $\tan^2\beta$ factors. 
Only the $\Phi_H t\bar{t}$ production reaches cross sections at the level 
of $\sim 0.3$~pb. For the detection, the $\gamma\gamma$ and the $b\bar{b}$ 
final states can be exploited. The latter being more promising due to 
slightly enhanced branching ratios compared to the SM case, 
cf.~Fig.~\ref{bran}. The vector boson fusion processes are two to 
three orders of 
magnitude smaller than the $gg\to \Phi$ and $b\bar{b}\Phi$ processes, 
reaching $\sim 5$~pb only for the SM--like Higgs boson with maximal coupling 
to the gauge bosons. In this case the $\tau^+\tau^-$ final states would allow 
for the detection by taking advantage of the energetic quark jets in the 
forward and backward directions \cite{tilman}.
The corresponding cross sections at the Tevatron show the same pattern being 
a factor 30 to 150 smaller depending on the process \cite{intense}. \\[0.2cm]
{\bf 5.2 Production at $e^+e^-$ colliders.}\smallskip \\
The main production processes at $e^+e^-$ colliders \cite{djouadi} 
are the bremsstrahlung process, $e^+e^- \to Z+h,H$, the 
associated production process, $e^+e^- \to A+h,H$, the vector boson fusion, 
$e^+e^- \to \bar{\nu}_e \nu_e / e^+e^- + h,H$, and the radiation off top and 
$b$ quarks, respectively, $e^+e^-\to t\bar{t}/ b\bar{b} + \Phi$, 
cf.~Fig.~\ref{eeprod}. Since 
the cross sections for Higgs-strahlung and Higgs pair production as well as 
the cross sections for the $h$ and $H$ production are mutually complementary 
to each other, and in view of the high luminosity that can be reached at an 
$e^+e^-$ collider \cite{ee} as well as the efficient $b$-quark tagging, all 
neutral Higgs bosons can be discovered at $e^+e^-$ machines in the 
intense--coupling regime. Furthermore, in the associated production 
processes with quark pairs, the $t\bar{t}\Phi_H$ and for large enough 
$\tan\beta$, the $b\bar{b}A,$ $b\bar{b}\Phi_A$ Yukawa couplings can be 
measured.
\begin{figure}[htbp]
\begin{center}
\vspace{0cm}
\epsfig{figure=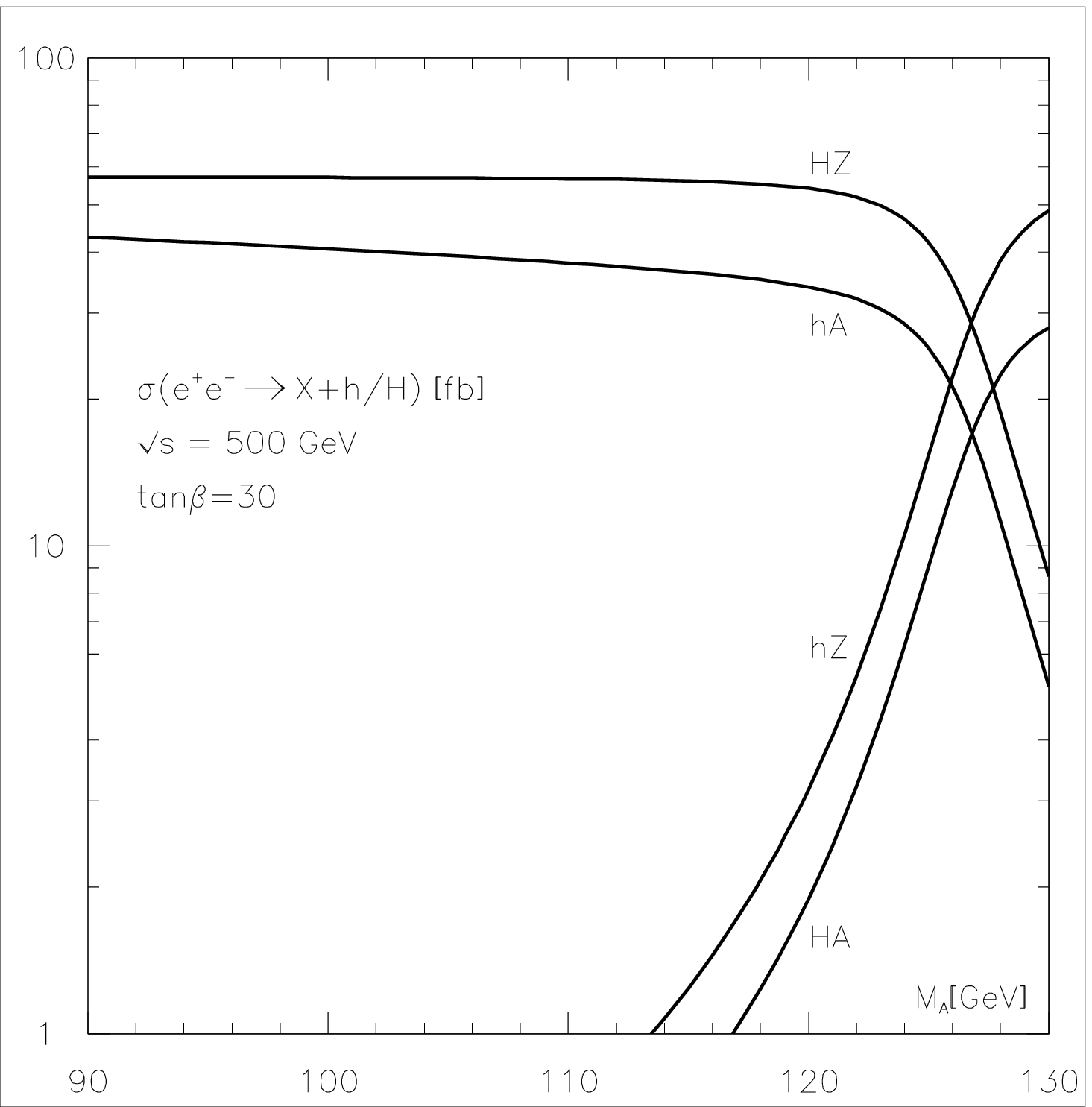,bbllx=20,bblly=26,bburx=410,bbury=408,height=4.2cm,width=4.2cm,clip=}\hspace{0.6cm}
\epsfig{figure=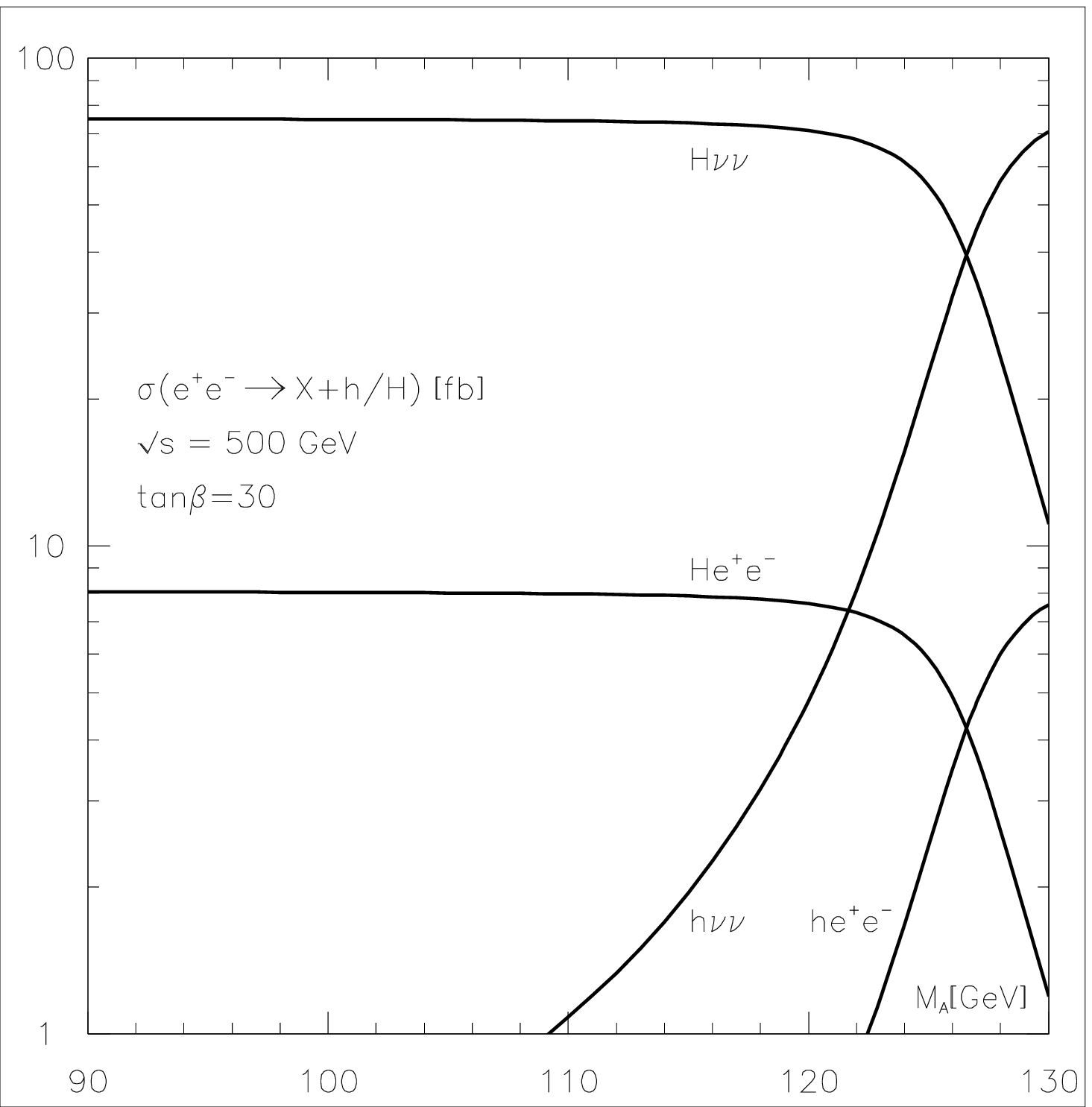,bbllx=20,bblly=26,bburx=410,bbury=408,height=4.2cm,width=4.2cm,clip=}\hspace{0.6cm}
\epsfig{figure=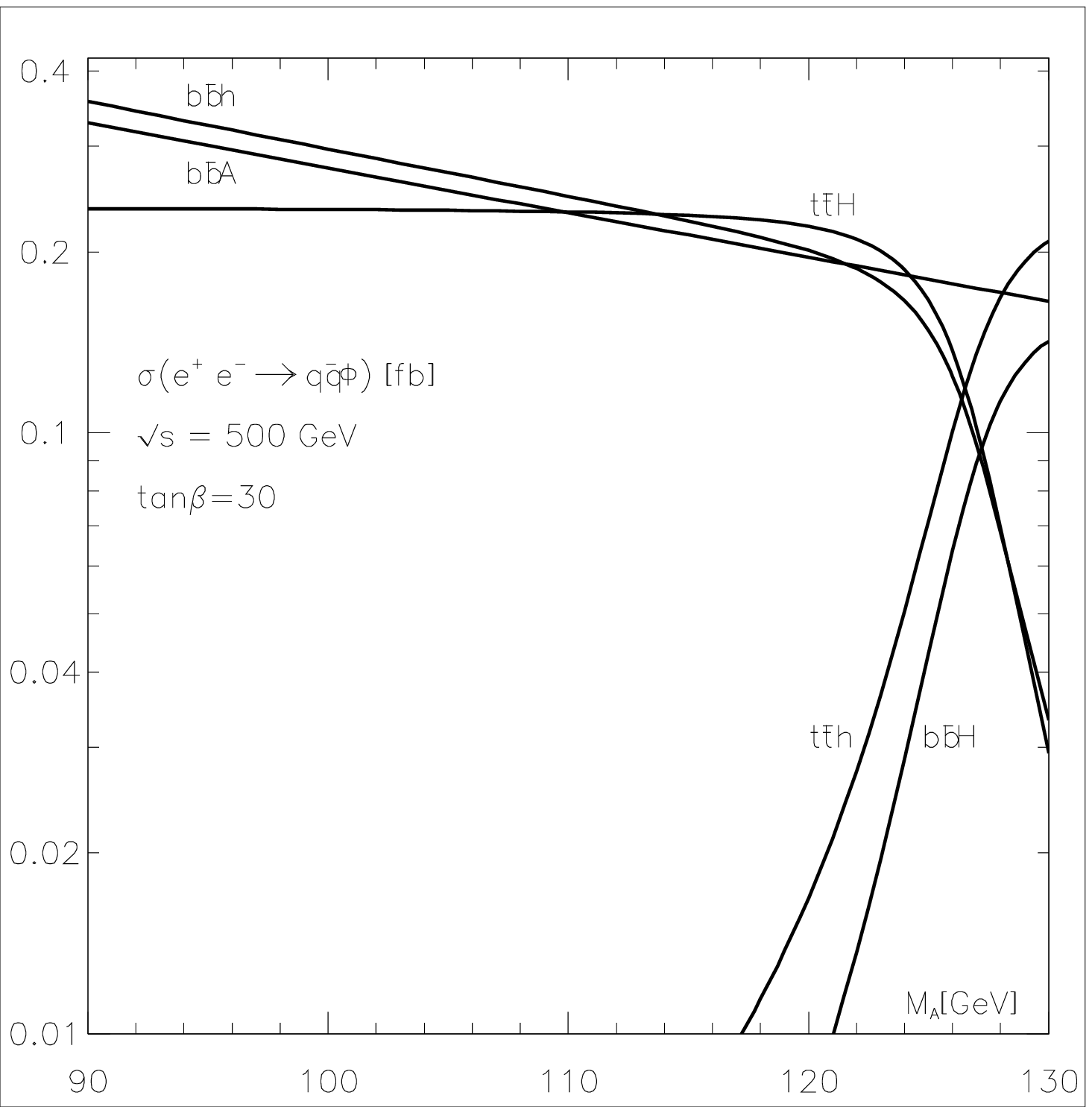,bbllx=20,bblly=26,bburx=410,bbury=408,height=4.2cm,width=4.2cm,clip=}
\end{center}
\vspace*{-.4cm}
\caption{{\it Higgs production cross sections at a 500~GeV $e^+e^-$ linear collider in the bremsstrahlung and associated Higgs production processes (left), the vector boson fusion processes (middle) and the associated production with top and bottom quarks (right) as a function of $M_A$ for $\tan\beta=30$. The SUSY parameters have been chosen as in Fig.~\ref{bran}.}}
\label{eeprod}
\vspace*{-.2cm}
\end{figure}\\[0.3cm]
%
%
%
%
\nn {\bf\large 6 Summary}
\\[0.2cm]
The preceding discussion has demonstrated that the phenomenology of the MSSM 
Higgs bosons in the intense--coupling regime is extremely rich. Being rather 
light, all Higgs bosons will be accessible at future colliders in various and 
complementary production channels. New features occurring in this scenario 
demand in some cases other techniques for the search of these particles than 
in the SM case and in the MSSM case close to the decoupling limit. Since the 
Higgs bosons are close in mass, it might be difficult to separate them, and
several production channels have to considered at the same time. In addition,
the large total widths for the pseudoscalar and pseudoscalar--like Higgs boson 
lead to broader signals. Finally, the clean $\gamma\gamma$ final state 
signatures used for the Higgs boson searches at the LHC, might be 
much less frequent than in the SM case. Having summarized the main features 
of the intense--coupling regime in this note, more detailed Monte-Carlo studies
are needed to assess at which extent these Higgs particles can be 
discovered and their properties can be measured at future colliders.
\bigskip \\
%
%
%
%
\nn {\bf\large Acknowledgments}
\\[0.2cm]
I would like to thank my collaborators E.~Boos, A.~Djouadi and A.~Vologdin 
for the fruitful collaboration on this project. This work has been supported
by the European Union under contract HPRN-CT-2000-00149.
%
%
%
%


\begin{thebibliography}{99} 

\bibitem{intense} For details see: E.~Boos, A.~Djouadi, M.~M\"uhlleitner and 
A.~Vologdin, Phys.~Rev.~D66 (2002) 055004, [hep-ph/0205160].
%
\bibitem{MSSM} For a review on the Higgs sector of the MSSM, see J.F. Gunion,
H.E. Haber, G.L. Kane and S. Dawson, ``The Higgs Hunter's Guide",
Addison--Wesley, Reading 1990.
%
\bibitem{RC} See {\it e.g.}~M.~Carena and H.E.~Haber, hep-ph/0208209, and 
references therein.

\bibitem{decoup} See {\it e.g.}~H.~Haber, hep-ph/9505240 and more recently 
A.~Dobado, M.J.~Herrero and S.~Penaranda, Eur.~Phys.~J.~C17 (2000) 487.

\bibitem{run2} S.~Abel et al., Report of the SUGRA working group for 
RUN II of the Tevatron, hep-ph/0003154.

\bibitem{cavalli} D.~Cavalli et al., Higgs report for the Les Houches 
2001 Workshop "Physics at TeV colliders", hep-ph/0203056. 

\bibitem{lhc} ATLAS Collaboration, Technical Design Report CERN-LHCC 99-14; 
CMS Collaboration, Technical Proposal, Report CERN-LHCC 94-38.

\bibitem{ee} E.~Accomando, Phys. Rept. 299 (1998) 1; TESLA, Technical Design 
Report, hep-ph/0106315 and hep-ex/0108012; American Linear Collider
Working Group, hep-ex/0106056; ACFA Linear Collider Working Group, 
hep-ph/0109166.

\bibitem{smlim} The LEP Higgs working group, hep-ex/0107029 and 
hep-ex/0107030; DELPHI Coll.~(P.~Abreu et al.), Phys.~Lett.~B499 (2001) 23; 
OPAL Coll.~(G.~Abbiendi et al.), Phys. Lett. B499 (2001) 38.

\bibitem{prec} Particle Data Group (D.E.~Groom et al.), Eur.~Phys.~J.~C15 
(2000) 1; The LEP Coll.s, the LEP EWWG and the SLD Heavy Flavor and EWWG, 
hep-ex/0103048; H.N.~Brown et al.~(muon ($g-2$) Collaboration), 
Phys.~Rev.~Lett.~86 (2001) 2227.

\bibitem{caretal} M.~Carena, S.~Mrenna and C.E.M.~Wagner, Phys.~Rev.~D62
(2000) 055008.

\bibitem{spira} M.~Spira et al.,  Nucl.~Phys.~B453 (1995) 17; 
A.~Djouadi and M.~Spira, Phys.~Rev.~D62 (2000) 014004.

\bibitem{tilman} T.~Plehn, D~Rainwater and D.~Zeppenfeld, 
Phys.~Rev.~D61 (2000) 093005.

\bibitem{djouadi} A.~Djouadi, J.~Kalinowski and P.M.~Zerwas, Z.~Phys.~C54 
(1992) 255 and Z.~Phys.~C57 (1993) 569. 

\end{thebibliography}
\end{document}